\documentclass[twocolumn,aps,prb]{revtex4}

\usepackage{amsmath,amssymb,amsfonts,bm}
\usepackage{graphicx,epstopdf}  %epsfig,
\usepackage{color}
\usepackage{lscape}
\usepackage{extarrows}
\usepackage{enumitem}
\usepackage{empheq}

\renewcommand{\d}{{\rm d}}

\newcommand{\w}{\omega}
\newcommand{\wti}{\widetilde}

\newcommand{\AB}{\mbox{\tiny $AB$}}
\newcommand{\BA}{\mbox{\tiny $BA$}}
\newcommand{\B}{\mbox{\tiny B}}
\newcommand{\s}{\mbox{\tiny S}}

\newcommand{\tS}{\mbox{\tiny S}}
\newcommand{\T}{\mbox{\tiny T}}

\newcommand{\SB}{\mbox{\tiny SB}}

\newcommand{\la}{\langle}
\newcommand{\ra}{\rangle}

\newcommand{\La}{\big\la}
\newcommand{\Ra}{\big\ra}

\newcommand{\Sec}[1]{Sec.\,\ref{#1}}
\newcommand{\nl}{\nonumber \\}
\newcommand{\be}{\begin{equation}}
\newcommand{\ee}{\end{equation}}
\newcommand{\bsube}{\begin{subequations}}
\newcommand{\esube}{\end{subequations}}
\newcommand{\Eq}[1]{Eq.\,(\ref{#1})}
\newcommand{\Eqs}[1]{Eqs.\,(\ref{#1})}
\newcommand{\Fig}[1]{Fig.\,\ref{#1}}

\newcommand{\RN}[1]{%
  \textup{\uppercase\expandafter{\romannumeral#1}}%
}

\begin{document}

\title{System--bath entanglement theorem with Gaussian 
environment{s}
}
\author{Peng-Li Du} \thanks{Authors of equal contributions}
\author{Yao Wang} \thanks{Authors of equal contributions}
\author{Rui-Xue Xu}\email{rxxu@ustc.edu.cn}
\author{Hou-Dao Zhang}
\author{YiJing Yan}

\affiliation{Hefei National Laboratory for Physical Sciences at the Microscale
and Department of Chemical Physics
and Synergetic Innovation Center of Quantum Information and Quantum Physics
and
Collaborative Innovation Center of Chemistry for Energy Materials (i{\rm ChEM}),
University of Science and Technology of China, Hefei, Anhui 230026, China}
\date{\today}
\begin{abstract}

In this work, we establish a so--called
``{\it system--bath entanglement theorem}'',
for arbitrary systems coupled with Gaussian environments.
This theorem connects the entangled system--bath response
functions in the total composite space
to those of local systems,
{ as long as the
interacting bath spectral densities are given.
}%
We validate the theorem with the direct evaluation via
the exact dissipaton--equation--of--motion approach.
Therefore, this work enables various quantum dissipation theories,
which originally describe only the reduced system dynamics, for their
evaluations on
the system--bath entanglement properties.
Numerical demonstrations are carried out on the Fano interference spectroscopies of spin--boson systems.

\end{abstract}
\maketitle

\section{Introduction}
\label{sec1}

System--bath entanglement plays a crucial role
in dynamic and thermal properties of complex systems.
However, most quantum dissipation theories (QDTs)
focus explicitly only on reduced system density operators.
This compromises the capabilities of conventional QDTs in evaluation on
such as the Fano resonances\cite{Fan611866,Mir102257,Lim17543,Zha09820,Tan1032}
and the correlated dynamics
between chromophores and surface plasmons.\cite{Luk10707,Fro125989,Rod15165}
System--bath entanglements also involve in
thermodynamics functions and thermal transports.

Almost all existing QDTs are based on Gaussian bath statistics.
Exact methods include the Feynman--Vernon influence
functional path integral formalism,\cite{Fey63118}
and its derivative equivalence,
the hierarchical--equations--of--motion (HEOM)
implementation.\cite{Tan906676,Tan06082001,Yan04216,Xu05041103} %,Xu07031107
Approximate methods often refer to quantum master
equations.\cite{Red651,Ber71539,Haa7398,Lin76393,Cal83587,%
Kam92,Yan982721,Yan002068,Xu029196,Zha987763,Jan08101104}
These include the Redfield theory and its modifications,\cite{Zha987763}
polaron--transformed versions,\cite{Jan08101104}
and self-consistent Born approximation
improvements.\cite{Lai914391,Jan022705,Esp09205303,Mav14054112,Jin14244111}
The simplicity of Gaussian environment is rooted at the
underlying Gauss--Wick's thermodynamics theorem.\cite{Wei08,Kle09,Yan05187}
The influence of environment
can be completely described within the linear response theory framework
in the isolated bare--bath subspace.
This feature has been exploited in various QDTs.

In this work, we address a missing ingredient, the so--called
``{\it system--bath entanglement theorem}'',
for an arbitrary system coupled with Gaussian environment.
As usual the total system--plus--bath composite Hamiltonian reads
\be\label{Hall}
  H_{\T}=H_{\s}+h_{\B}+H_{\SB}\equiv H_{\s}+h_{\B}+\sum_a\hat Q_a\hat F_{a}.
\ee
The system Hamiltonian $H_{\s}$ and dissipative modes $\{\hat Q_a\}$ are arbitrary.
In the above equation, we denote the bath Hamiltonian
in lower case for the Gaussian environment scenario.
This requires not only $h_{\B}$ be harmonic but also
the hybrid bath modes $\{\hat F_a\}$ be linear.
That is
\be\label{gwmodel}
 h_{\B}=\frac{1}{2}\sum_j\omega_j(\hat p_j^2+\hat x_j^2) \ \ {\rm and}\  \
 \hat F_{a}=\sum_j c_{aj}\hat x_j.
\ee
{
These microscopic expressions, with
dimensionless coordinates $\{\hat x_j\}$ and momentums $\{\hat p_j\}$,
will be used explicitly later in \Sec{sec_theoA}.
}
%%%
Throughout the paper we set
$\hbar=1$ and
$\beta=1/(k_BT)$, with $k_B$ the Boltzmann constant and $T$ the temperature.

{
 It is worth noting that open quantum system
is subject to dephasing, energy relaxation, and transport.
These irreducible processes are beyond the total composite Hamiltonian
description.
%%%
Additional information, such as temperature $T$
and the interacting bath statistics, would be
needed.\cite{Wei08,Kle09,Yan05187}
%%%%
In fact, the total composite Hamiltonian, $H_{\T}$ of \Eq{Hall},
constitutes a ``closed system'' in the thermodynamics nomenclature,
which is in thermal contact with surroundings at a given temperature $T$.
%%%
The resultant thermal equilibrium is given by
the total system--and--bath composite density operator,
$\rho^{\rm eq}_{\T}(T)=e^{-\beta H_{\T}}/Z{\T}$,
with $Z_{\T}\equiv {\rm Tr}e^{-\beta H_{\T}}$ 
being the thermodynamics partition functions.
One typical example is a total composite solution system in chemistry,
where $H_{\tS}$ and $h_{\B}$ stand for the solute particle and solvent
environment, respectively,
and $H_{\SB}$ for the coupling between them.
%%%
Apparently, physically relevant and directly measurable quantities
such as correlation and response functions
are all concerned with the total system--plus--bath composite space.
%%%

 The simplicity of  Gaussian environment, \Eq{gwmodel},
is that its influence on the reduced system
can be completely described with
the linear response theory in the bare--bath subspace.
\cite{Wei08,Kle09,Yan05187}
The fundamental quantities here are the
interacting bath response functions,
}
\be\label{phidef}
 \phi_{ab}(t-\tau)\equiv
     i\la[\hat F^{\B}_{a}(t),\hat F^{\B}_b(\tau)]\ra_{\B},
\ee
where
$\hat F^{\B}_{a}(t)\equiv e^{ih_{\B}t}\hat F^{\B}_{a}e^{-ih_{\B}t}$
and
 $\la(\,\cdot\,)\ra_{\B}\equiv{\rm tr}_{\B}[(\,\cdot\,)\rho^{0}_{\B}(T)]$
with
 $\rho^{0}_{\B}(T)\equiv e^{-\beta h_{\B}}/{\rm tr}_{\B}(e^{-\beta h_{\B}})$.
%%%
{ 
These are just uncorrelated bare--bath quantities, as 
if $H_{\s\B}=0$.
The interacting bath spectral density functions
are \cite{Wei08,Kle09,Yan05187}
\be\label{Jab_w}
 J_{ab}(\w)=\frac{1}{2i}\int^{\infty}_{-\infty}\!\!\d t\,
  e^{i\w t}\phi_{ab}(t),
\ee
which are often given through models in various
QDTs.\cite{Tan06082001,Wei08,Kle09,Yan05187,Leg871,Mei993365,Tho012991}
%%%
However, most of these theories focus only on the reduced system
dynamics and evaluate expectation values
and correlation/response functions
of various system operators $\{\hat O_{\s}\}$.

 As mentioned earlier, the total composite Hamiltonian
$H_{\T}$ of \Eq{Hall} constitutes a closed system in thermodynamics.
It is important to include the system--bath entanglement dynamics
arising from $H_{\SB}=\sum_a\hat Q_a\hat F_{a}$, the last term in \Eq{Hall},
into explicit consideration.
In fact, $H_{\SB}$ is related to the system--bath hybridization
free--energy change.\cite{Kir35300}
It would  require the response/correlation functions
between system operators $\{\hat Q_{a}\}$
and the hybrid bath modes $\{\hat F_a\}$ in the total system--and--bath
composite space.

 The system--bath entanglement theorem to be established
in this work relates the $i\la[\hat Q_{a}(t),\hat F_b(0)]\ra$
and $i\la[\hat F_{a}(t),\hat F_b(0)]\ra$
types of response functions
to those of local system $i\la[\hat Q_{a}(t),\hat Q_b(0)]\ra$.
Here, $\hat O(t)\equiv e^{iH_{\T}t}\hat Oe^{-iH_{\T}t}$
and $\la(\,\cdot\,)\ra\equiv{\rm Tr}[(\,\cdot\,)\rho^{\rm eq}_{\T}(T)]$,
defined in the total system--plus--bath composite space.
%%%
Apparently, $i\la[\hat F_{a}(t),\hat F_b(0)]\ra\neq \phi_{ab}(t)$
of \Eq{phidef}.
The latter is the uncorrelated bare--bath subspace property,
which will serve as the bridge to the aforementioned
relations.
%%%%
 The conventional QDTs,  such as the HEOM formalism,\cite{Tan906676,Tan06082001,Yan04216,Xu05041103}
%,Xu07031107
are capable of evaluating the local system properties.
}%%%%%%%
This work would naturally enable their evaluations on those
entanglement response/correlation functions
between local system and non-local
environment.
These are all the ingredients in Fano interference
spectroscopies.\cite{Fan611866,Mir102257,Lim17543,Zha15024112,Xu151816}
%%%

 It is worth noting that we will establish
the system--bath entanglement theorem
in non-equilibrium steady--state scenario.
Therefore, this work would
be closely related to plasmon
spectroscopies dressed with strong plasmonic fields.\cite{Luk10707,Fro125989,Rod15165}
Other methods such as non-equilibrium Green's function technique
would also be
{ enabled for the aforementioned}
system--bath entanglement properties.
%%%
{
Moreover, one may exploit the system--bath entanglement theorem
to bridge between all--atom simulations and implicit Gaussian
solvent environment models.

}

This paper is organized as follows.
We establish the system--bath entanglement theorem in \Sec{sec_theo}
and numerically demonstrate it in \Sec{sec_num}.
Validations are carried out with respect to direct evaluation via
the exact dissipaton--equation--of--motion (DEOM) approach.\cite{Yan14054105,Yan16110306}
Fano interference spectroscopies are evaluated
on spin--boson systems.
We conclude this work in \Sec{sec_sum}.

\section{System--bath entanglement theorem}
\label{sec_theo}
\subsection{Langevin equation for solvation dynamics}
\label{sec_theoA}

Consider the quantum Langevin equation
for the hybrid bath dynamics,
as implied in the total system--and--bath composite Hamiltonian, $H_{\T}$
of \Eq{Hall} with \Eq{gwmodel}.
Let $\hat O(t)\equiv e^{iH_{\T}t}\hat Oe^{-iH_{\T}t}$.
We obtain
\be\label{eq05}
\ddot{\hat x}_j(t)=-\omega^2_j\hat x_j(t)-\omega_j\sum_a c_{aj}\hat Q_a(t).
\ee
Its formal solution is
\begin{align}\label{eq06}
   \hat x_j(t)&=\hat x_j(0)\cos(\omega_jt)+\hat p_j(0)\sin(\omega_jt)
\nl & \qquad -\sum_a c_{aj}
   \int_0^t\!\!{\rm d}\tau\,\sin[\omega_j(t-\tau)]\hat Q_a(\tau).
\end{align}
It together with the second identity of \Eq{gwmodel} lead to
\be\label{eq07}
   \hat F_{a}(t)=\hat F^{\B}_{a}(t)-\sum_b\int_0^t\!\!{\rm d}\tau\,\phi_{ab}(t-\tau)\hat Q_b(\tau),
\ee
with the bare--bath random force
{ operator,
$\hat F^{\B}_{a}(t)\equiv e^{ih_{\B}t}\hat F^{\B}_{a} e^{-ih_{\B}t}$ that
also involves in $\phi_{ab}(t)$ of \Eq{phidef},
the expression,
}
\begin{align}\label{eq08}
  \hat F^{\B}_{a}(t) = \sum_j c_{aj}[\hat x_j(0)\cos(\omega_jt)+\hat p_j(0)\sin(\omega_jt)].
\end{align}
It is easy to obtain
\be\label{eq09}
  i[\hat F^{\B}_{a}(t),\hat F_b(0)]=
  i[\hat F^{\B}_{a}(t),\hat F^{\B}_b(0)]=
 \phi_{ab}(t).
%  i\la[\hat F^{\B}_{a}(t),\hat F^{\B}_b(\tau)]\ra_{\B}
   % =\tfrac{1}{\hbar}\sum_j c_{aj}c_{bj}\sin(\omega_jt).
\ee
This commutator itself is a c-number and equals to the
bare--bath response function, \Eq{phidef}.

Equation (\ref{eq07}) describes the Langevin dynamics for the hybridizing bath modes.
It differs from traditional Langevin equations
which focus on reduced systems.
However this serves as the starting point to
the following establishment of system--bath entanglement theorem.
Let $\chi_{\AB}(t-\tau)\equiv i\la[\hat A(t),\hat B(\tau)]\ra$
be the response function in the total composite space.
%for two arbitrary Hermite operators.
%Apparently, $\chi_{\AB}(t)=-\chi_{\BA}(-t)$, as a general antisymmetric relation for response functions.
As inferred from \Eq{eq08}, $[\hat F^{\B}_{a}(t),\hat O_{\s}]=0$ for
an arbitrary system operator $\hat O_{\s}$.
Consequently, \Eq{eq07} results in
\be\label{eq10}
  \la [\hat F_a(t),\hat O_{\s}(0)]\ra
    =-\sum_b\!\int_0^t\!\!{\rm d}\tau\,\phi_{ab}(t-\tau)
    \la [\hat Q_b(\tau),\hat O_{\s}(0)]\ra.
\ee
This expresses the non--local response as the convolution between
the bare--bath and the local--system properties.

\subsection{System--bath entanglement theorem}

In relation to the system--and--bath
entanglement dynamics underlying the hybridization
system and bath modes, $\{\hat Q_a\}$ and $\{\hat F_a\}$,
%These are the operators that participate in
%the coupling $H_{\tS\B}$.
%%%
{ denote in the following
response functions in the total composite space,
\begin{align}\label{eq11}
\begin{split}
  \chi^{\s\s}_{ab}(t)&\equiv i\la[\hat{Q}_a(t),\hat{Q}_b(0)]\ra ,
\\
  \chi^{\s\B}_{ab}(t)&\equiv i\la[\hat{Q}_a(t),\hat{F}_b(0)]\ra,
\\
  \chi^{\B\s}_{ab}(t)&\equiv i\la[\hat{F}_a(t),\hat{Q}_b(0)]\ra,
\\
  \chi^{\B\B}_{ab}(t)&\equiv i\la[\hat{F}_a(t),\hat{F}_b(0)]\ra.
\end{split}
\end{align}
The involving operators all arise from $H_{\SB}=\sum_a\hat Q_a\hat F_{a}$,
with specifying $\{\hat Q_a\}$ and $\{\hat F_{a}\}$
being the operators in the system and bath subspaces,
respectively.
}
Equation (\ref{eq07}) gives rise to
\be\label{eq12}
  \chi^{\B\tS}_{ab}(t)=-\sum_{b'}\!\int_0^t\!\!{\rm d}\tau\,
   \phi_{ab'}(t-\tau)\chi^{\tS\tS}_{b'b}(\tau)
\ee
By $\chi^{\tS\B}_{ba}(t)=-\chi^{\B\tS}_{ab}(-t)$, it leads to further
\begin{align}\label{eq13}
  \chi^{\tS\B}_{ba}(t)
%&=\sum_{b'}\!\int_0^{-t}\!\!\!{\rm d}\tau\,\phi_{ab'}(-t-\tau)
 %   \chi^{\tS\tS}_{b'b}(\tau)
%%
%\nl&
 =-\sum_{b'}\!\int_0^t\!\!{\rm d}\tau\,
    \phi_{b'a}(t-\tau)\chi^{\tS\tS}_{bb'}(\tau),
\end{align}
obtained via changing the integral variable
with $\tau'=-\tau$, followed by using
the antisymmetric relation
for both response functions in the integrand.
Moreover, \Eq{eq07} also gives rise to
\be\label{eq14}
 \chi^{\B\B}_{ab}(t) = \phi_{ab}(t)
- \sum_{b'}\!\int^{t}_{0}\!\d\tau\,\phi_{ab'}(t-\tau)
  \chi^{\tS\B}_{b'b}(\tau)
\ee

In terms of the susceptibility or frequency resolution,
$\wti f(\omega)=\int_0^\infty {\rm d}t\,e^{i\omega t}f(t)$,
\Eqs{eq12}--(\ref{eq14}) read
\begin{align}
\begin{split}
  \wti\chi^{\B\tS}_{ab}(\omega)
&=-\sum_{b'}\wti\phi_{ab'}(\w)\wti\chi^{\tS\tS}_{b'b}(\w)
\\
  \wti\chi^{\tS\B}_{ba}(\omega)
&=-\sum_{b'}\wti\chi^{\tS\tS}_{bb'}(\w)\wti\phi_{b'a}(\w)
\end{split}
\label{eq15}
%%%
\intertext{and}
%%%
 \wti\chi^{\B\B}_{ab}(\w)
&=\wti\phi_{ab}(\w)+\sum_{a'b'}\wti\phi_{aa'}(\w)
   \wti\chi^{\tS\tS}_{a'b'}(\w)\wti\phi_{b'b}(\w)
\label{eq16}
\end{align}
 Let $\wti{\bm\chi}^{\B\tS}(\w)\equiv \{\wti\chi^{\B\tS}_{ab}(\w)\}$
be a matrix, and similar for others, so that
one can recast \Eqs{eq15} and (\ref{eq16}) as
\begin{align}
  \wti{\bm\chi}^{\B\tS}(\w)&=-\wti{\bm\phi}(\w)\wti{\bm\chi}^{\tS\tS}(\w),
\
  \wti{\bm\chi}^{\tS\B}(\w)=-\wti{\bm\chi}^{\tS\tS}(\w)\wti{\bm\phi}(\w),
\label{eq17}
\intertext{and}
  &\wti{\bm\chi}^{\B\B}(\w) = \wti{\bm\phi}(\w)
  +\wti{\bm\phi}(\w)\wti{\bm\chi}^{\tS\tS}(\w)\wti{\bm\phi}(\w).
\label{eq18}
\end{align}
We refer these identities the system--bath entanglement theorem
that goes with the Gaussian bath model.
They relate the nonlocal properties,
$\wti{\bm\chi}^{\B\tS}(\w)$, $\wti{\bm\chi}^{\tS\B}(\w)$ and
$\wti{\bm\chi}^{\B\B}(\w)$,
with the local system $\wti{\bm\chi}^{\tS\tS}(\w)$
and the bare bath $\wti{\bm\phi}(\w)$.
Define the overall system--bath entanglement susceptibility,
\be\label{eq19}
\begin{split}
  {\chi}_{\tS\B}(\w)&\equiv \sum_{a}\wti{\chi}^{\tS\B}_{aa}(\w)
= {\rm tr}\wti{\bm\chi}^{\tS\B}(\w),
\\
 {\chi}_{\B\tS}(\w) &\equiv \sum_{a}\wti{\chi}^{\B\tS}_{aa}(\w)
= {\rm tr}\wti{\bm\chi}^{\B\tS}(\w).
\end{split}
\ee
From \Eq{eq17} we have immediately
\be\label{eq20}
  {\chi}_{\tS\B}(\w)={\chi}_{\B\tS}(\w).
%%%% =-{\rm tr}[\wti{\bm\phi}(\w)\wti{\bm\chi}^{\tS\tS}(\w)]
\ee
This describes the reciprocal relation of the overall
system--bath entanglement susceptibility.

It is worth emphasizing that the ensemble averages underlying all
 response functions in this work are concerned with
steady--state scenario.
In other words, the established theorem,
from \Eq{eq10} to \Eq{eq20}, is for arbitrary systems
coupled with Gaussian steady--state environments.

It is also noticed that
in general, the frequency resolution can be expressed as
$\wti\chi_{\AB}(\omega)=\wti\chi_{\AB}^{(+)}(\omega)
 +i\wti\chi_{\AB}^{(-)}(\omega)$,
with $\wti\chi_{\AB}^{(\pm)}(\omega)=[\wti\chi_{\BA}^{(\pm)}(\omega)]^\ast$
being Hermite/anti-Hermite matrix component, respectively.
Here $\hat A$ and $\hat B$ are both Hermitian operators.
The anti-Hermite component, $\wti\chi_{\AB}^{(-)}(\omega)$,
refers also the spectral density.
In the thermal equilibrium scenario, it is related to the correlation function via the
fluctuation--dissipation theorem.\cite{Wei08,Kle09,Yan05187}

\section{Numerical demonstrations}
\label{sec_num}

\subsection{Fano profile in a spin--boson model}

{ For numerical
demonstrations, we focus on
a single--dissipative mode spin--boson model
and evaluate the Fano interference spectroscopy.
We will see that the system--bath entanglement
theorem--based indirect evaluations
perfectly agree with the results of direct approach.\cite{Zha15024112}
%%%

The total system--and--bath composite model Hamiltonian,
in the presence of an external field $E(t)$,
assumes
}
\be\label{eq21}
  H_{\T}(t)=\frac{\Omega}{2}\hat\sigma_z+h_{\B}
   +\hat\sigma_x\hat F-\hat\mu_{\T}E(t),
\ee
where
\be\label{eq22}
   \hat\mu_{\T}=\mu_{\s}\hat\sigma_x+\nu_{\B}\hat F.
\ee
%%%
{
The first term, $\hat\mu_{\tS}=\mu_{\s}\hat\sigma_x$,
represents
the transition dipole of the two--level
system (or solute molecule),
which itself has no permanent dipole.
The second term in \Eq{eq22} describes the
external light field $E(t)$--induced bath (solvent)
environment polarization.
%%%
Physically, this would correspond to the scenario,
where individual solvent molecule has small polar, but with random
orientation. Thus, the bulk of solvent is isotropic around
the nonpolar solute molecule.
%%%
The external light field, assumed to be linear polarized,
breaks the original isotropic symmetry
and induces the solvent polarization.
On the other hand, due to
the form of system--bath coupling in \Eq{eq21},
the transition dipole of the solute system
could also induce
the polarized solvation coordinate $\hat F$.
%%%

\begin{figure}
\includegraphics[width=0.43\textwidth]{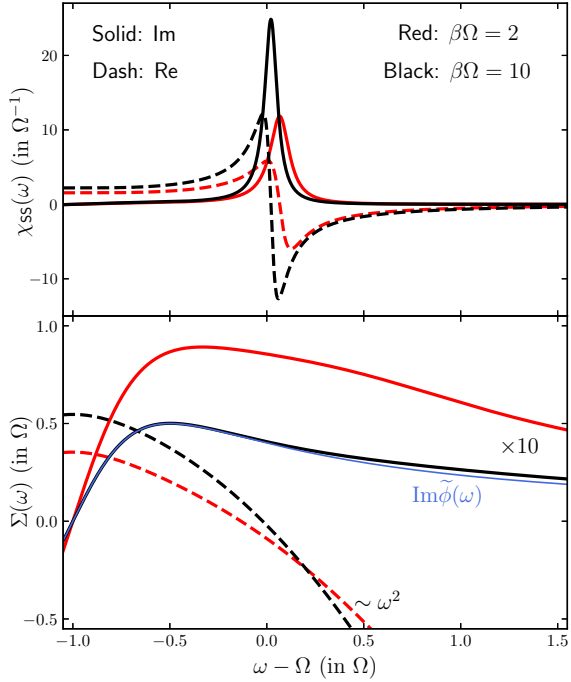}
\caption{Real (dash) and imaginary (solid) parts of
 $\chi_{\s\s}(\w)$ (upper--panel) and $\Sigma(\w)$ (lower--panel),
 at a low temperature ($\beta\Omega=10$; black)
 and a high temperature ($\beta\Omega=2$; red).
 Depicted is also ${\rm Im}\,\wti\phi(\w)$ (thin-solid) for comparison.
 The three solid curves in the lower panel are timed with 10.
 The asymptotic behavior of ${\rm Re}\,\Sigma(\w)$ 
 to infinity is found to be quadratic.
 Drude spectral bath model 
 $\wti\phi(\w)=2\lambda\gamma/(\gamma-i\w)$  is adopted
 with the parameters chosen as 
 $\lambda=0.05\Omega$ and $\gamma=0.5\Omega$.
}\label{fig1}
\end{figure}

 Nevertheless, for the main purpose of the present work,
we set the bath polarization,
the second term in \Eq{eq22},
for the form of $\hat\mu_{\B}=\nu_{\B}\hat F$.
}
The total composite  dipole susceptibility is then given by
\begin{align}\label{eq23}
{ \Xi(\omega)}
&\equiv i\!\int_0^\infty\!\!{\rm d}t\,e^{i\omega t}
 \La[\hat\mu_{\T}(t),\hat\mu_{\T}(0)]\Ra
\nl&=\mu_{\s}^2 \chi_{\s\s}(\w) +2\mu_{\s}\nu_{\B}\chi_{\s\B}(\w)
 +\nu_{\B}^2 \chi_{\B\B}(\w),
\end{align}
with [cf.\ \Eqs{eq11} and (\ref{eq20})]
\be\label{eq24}
\begin{split}
  \chi_{\s\s}(\w)&\equiv i\!\int_0^\infty\!\!{\rm d}t\,e^{i\omega t}
    \la[\hat\sigma_x(t),\hat\sigma_x(0)]\ra,
\\
  \chi_{\s\B}(\w)&\equiv i\!\int_0^\infty\!\!{\rm d}t\,e^{i\omega t}
   \la[\hat\sigma_x(t),\hat F(0)]\ra,
\\
  \chi_{\B\B}(\w)&\equiv i\!\int_0^\infty\!\!{\rm d}t\,e^{i\omega t}
   \la[\hat F(t),\hat F(0)]\ra.
\end{split}
\ee
The system--bath entanglement theorem, \Eqs{eq17} and (\ref{eq18}),
leads to \Eq{eq23} further the expression,
\be\label{eq25}
  \Xi(\w)=\nu_{\B}^2\wti\phi(\w)
  +[\mu_{\s}-\nu_{\B}\wti\phi(\w)]^2 \chi_{\s\s}(\w).
\ee
%%%
To proceed, we denote
\be\label{ea26}
\begin{split}
 &\phi_r(\w)\equiv {\rm Re}\,\wti\phi(\w),
\ \ \phi_i(\w)\equiv {\rm Im}\,\wti\phi(\w),
\\ &\qquad
 q(\w)\equiv \mu_{\s} -\nu_{\B}\phi_r(\w).
\end{split}
\ee
Introduce further
\be\label{ea27}
 {z}(\w) \equiv \frac{\chi_{\tS\tS}(\w)}{|\chi_{\s\s}(\w)|^2}
%\equiv \frac{
%  \wti\chi'_{\tS\tS}(\w)+i\wti\chi''_{\tS\tS}(\w)
%  }{|\wti\chi_{\s\s}(\w)|^2}
\equiv {z}_r(\w)+i{z}_i(\w).
\ee
Here, ${z}_r(\w)\equiv {\rm Re}\,{z}(\w)$ and
      ${z}_i(\w)\equiv {\rm Im}\,{z}(\w)$.
Some simple algebra leads to
\Eq{eq25} the expressions,
\begin{align}\label{eq28}
 \frac{{\rm Re}\,\Xi(\w)}{|\chi_{\s\s}(\w)|^2}
&= \frac{\nu_{\B}^2 \phi_r(\w)}{|\chi_{\s\s}(\w)|^2}
  +2\nu_{\B}\phi_i(\w)q(\w){z}_i(\w)
\nl &\quad
   +[q^2(\w)-\nu_{\B}^2\phi_i^{2}(\w)]{z}_r(\w),
\end{align}
and
\begin{align}\label{eq29}
  \frac{{\rm Im}\,\Xi(\w)}{|\chi_{\s\s}(\w)|^2 }
&=[{z}_i(\w)-\phi_i(\w)] [ q^2(\w) + {\nu^2_{\B}\phi_i(\w)}{z}_i(\w)]
\nl&\quad
  +\phi_i(\w)[q(\w)-\nu_{\B}{z}_r(\w)]^2.
\end{align}

\begin{figure}
\includegraphics[width=0.43\textwidth]{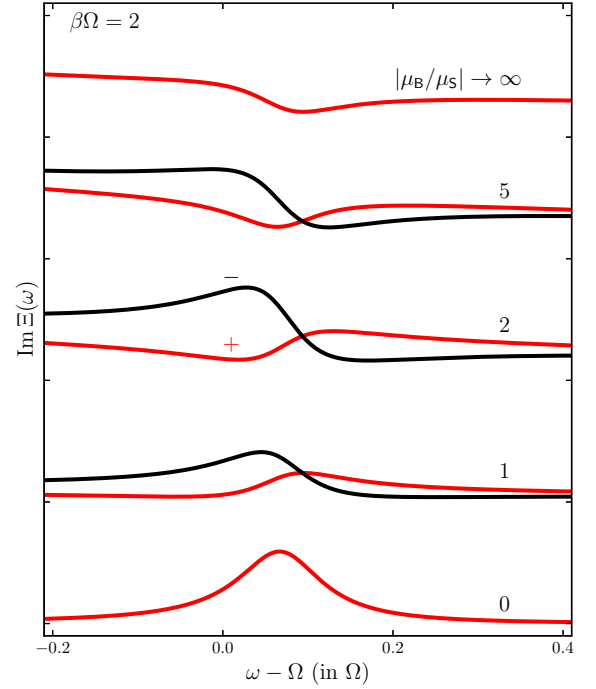}
\caption{Response spectra,
 ${\rm Im}\,\Xi(\w)$ [cf.\ \Eq{eq25}],
 for the high temperature case of \Fig{fig1}.
 The bath dipole $\mu_{\B}\equiv 2\lambda\nu_{\B}$
 is chosen as
 $\mu_{\B}/\mu_{\tS}=\infty,\pm 5, \pm 2, \pm 1, 0$.
 The red and black curves correspond to the $+$ and $-$ signs, 
 respectively.
}\label{fig2}
\end{figure}

In fact, $z(\w)$ of \Eq{ea27}
is related to the self--energy $\Sigma(\w)$ in
\be\label{eq30}
 \chi_{\s\s}(\w)=\frac{\Omega}{\Omega^2-\w^2-\Omega{\Sigma}(\w)},
\ee
via
\be\label{eq31}
\begin{split}
   &{\rm Re}\,{\Sigma}(\w)\equiv[\Omega^2-\w^2-\Omega{z}_r(\w)]/\Omega,
 \\
   &{\rm Im}\,{\Sigma}(\w)\equiv{z}_i(\w).
\end{split}
\ee
Note that the boson--boson model goes with
$\Sigma(\w)=\wti\phi(\w)$.\cite{Wei08,Kle09,Yan05187}
However, for spin--boson model, the self--energy needs to be evaluated
via $\chi_{\s\s}(\w)$  from certain QDT methods.

Figure 1 depicts $\chi_{\s\s}(\w)$ (upper--panel) and
the associated self--energy
$\Sigma(\w)$ (lower--panel), at a low temperature ($\beta\Omega=10$; black)
and a high temperature ($\beta\Omega=2$; red).
Here, $\Sigma(\w)$ is obtained via \Eq{eq30} from
$\chi_{\s\s}(\w)$.
The latter is evaluated via the exact DEOM approach,\cite{Yan14054105,Yan16110306}
which is equivalent to the HEOM method,\cite{Tan906676,Tan06082001,Yan04216,Xu05041103} %,Xu07031107
in the absence of bath polarization.
Physically, Im\,$\chi_{\s\s}(\w)$ and Re\,$\chi_{\s\s}(\w)$ are
related to the reduced system linear spectrum and dispersion, respectively,
in the absence of bath polarization $(\nu_{\B}=0)$.
As anticipated, the peak is
relatively sharp and strong in the low--temperature regime.
In the lower panel, the asymptotic behavior of 
${\rm Re}\,\Sigma(\w)$
is found to be quadratic in the large $\w$ regime.
This is also the behavior of ${z}_r(\w)$, cf.\ \Eq{eq31}.
As mentioned earlier, the boson--boson model goes with
$\Sigma(\w)=\wti\phi(\w)$.
We do observe that ${\rm Im}\,\Sigma(\w)\approx\phi_{i}(\w)$,
as anticipated for the low temperature case studied here,
cf.\ the black versus thin-blue curves in the lower panel.

Figure 2 exhibits the Fano
interference spectral lineshape, ${\rm Im}\,\Xi(\w)$,
with different values of relative bath dipole strength $\mu_{\B}/\mu_{\s}$, where $\mu_{\B}\equiv 2\lambda\nu_{\B}$.
It is noticed that the dipole ratio can be either positive or negative.
The plus and minus signs in the figure represent the
 two special cases where the bath dipole is parallel and anti-parallel to the system one, respectively.
As mentioned above, the differences between
${\rm Im}\,\Sigma(\w)$ %[equal to ${z}_i(\w)$; \Eq{eq31}]
and $\phi_{i}(\w)$
become smaller
as the temperature decreases.
This would lead to more similarities of Fano
interference patterns between the spin--boson and boson--boson cases,\cite{Zha15024112,Xu151816}
cf.\ \Eq{eq29}.
All results of $\Xi(\w)$ here are evaluated via \Eq{eq25}
from $\chi_{\s\s}(\w)$ and $\wti\phi(\w)\equiv \phi_{r}(\w)+i\phi_{i}(\w)$,
which have been confirmed to be consistent with those from
the direct DEOM evaluations on \Eq{eq23}.\cite{Zha15024112,Xu151816}
Thus the system--bath entanglement theorem [\Eqs{eq17} and (\ref{eq18})]
is also numerically verified.

\section{Summary}
\label{sec_sum}
In this work, we propose the {system--bath entanglement theorem},
\Eqs{eq17} and (\ref{eq18}),
for arbitrary systems coupled with Gaussian environments.
This theorem expresses the entangled system--bath response
functions in the total composite space
with those of local system,
{ as long as the
interacting bath response functions $\{\phi_{ab}(t)\}$
of \Eq{phidef} are given.
This is the general case as the
quantum dissipation formulations
are concerned with.
The captioned theorem}
is established on basis of the convolution relation, \Eq{eq10},
 between
the bare--bath and the local--system responses for
non--local system--bath properties,
obtained by revisiting the Langevin dynamics for the hybridizing bath modes,
\Eq{eq07}.
This theorem enables various quantum dissipation theories,
which originally only deal with the reduced system dynamics,
 to evaluate
system--bath entanglement properties.

To ``visualize'' the theorem, we evaluate the Fano interference spectra
of spin--boson systems via both direct DEOM approach on \Eq{eq23} and indirect entanglement--theorem approach on \Eq{eq25}.
We obtain full consistency between the results from these two approaches.
The Fano analysis made here, \Eqs{eq22}--(\ref{eq31}),
could be readily extended to more complex systems.
Noticed that the system--bath entanglement theorem here is established
in non-equilibrium steady--state scenario.
Therefore it is anticipated to
be closely related to plasmon
spectroscopies dressed with strong plasmonic fields.
Moreover, other methods such as non-equilibrium Green's function technique
would also be readily exploited for
system--bath entanglement properties.

\begin{acknowledgements}

 Support from
the Ministry of Science and Technology of China (Nos.\ 2017YFA0204904 \&
2016YFA0400904) and
the Natural Science Foundation of China (Nos.\ 21633006 \& 21703225)
is gratefully acknowledged.

\end{acknowledgements}

%\bibliographystyle{./aiptit}
%\bibliography{./bibrefs}

\begin{thebibliography}{10}

\bibitem{Fan611866}
U.~Fano, \newblock ``Effects of configuration interaction on intensities and
  phase shifts,'' Phys. Rev. {\bf 124}, 1866 (1961).

\bibitem{Mir102257}
A.~E. Miroshnichenko, S.~Flach, and Y.~S. Kivshar, \newblock ``Fano resonances
  in nanoscale structures,'' Rev. Mod. Phys. {\bf 82}, 2257 (2010).

\bibitem{Lim17543}
M.~F. Limonov, M.~V. Rybin, A.~N. Poddubny, and Y.~S. Kivshar, \newblock ``Fano
  resonances in photonics,'' Nature Photonics {\bf 11}, 543 (2017).

\bibitem{Zha09820}
Y.~Zhang, T.-T. Tang, C.~Girit, Z.~Hao, M.~C. Martin, A.~Zettl, M.~F. Crommie,
  Y.~R. Shen, and F.~Wang, \newblock ``Direct observation of a widely tunable
  bandgap in bilayer graphene,'' Nature {\bf 459}, 820 (2009).

\bibitem{Tan1032}
T.-T. Tang, Y.~Zhang, C.-H. Park, B.~Geng, C.~Girit, Z.~Hao, M.~C. Martin,
  A.~Zettl, M.~F. Crommie, S.~G. Louie, Y.~R. Shen, and F.~Wang, \newblock ``A
  tunable phonon-exciton Fano system in bilayer graphene,'' Nature Nanotech.
  {\bf 5}, 32 (2010).

\bibitem{Luk10707}
B.~Luk$'$yanchuk, N.~I. Zheludev, S.~A. Maier, N.~J. Halas, P.~Nordlander,
  H.~Giessen, and C.~T. Chong, \newblock ``The Fano resonance in plasmonic
  nanostructures and metamaterials,'' Nature Materials {\bf 9}, 707 (2010).

\bibitem{Fro125989}
R.~R. Frontiera, N.~L. Gruenke, and R.~P. {Van Duyne}, \newblock ``Fano-like
  resonances arising from long-lived molecule-plasmon interactions in colloidal
  nanoantennas,'' Nano Lett. {\bf 12}, 5989 (2012).

\bibitem{Rod15165}
D.~Rodrigo, O.~Limaj, D.~Janner, D.~Etezadi, F.~J. Garc{\'\i}a~de Abajo,
  V.~Pruneri, and H.~Altug, \newblock ``Mid-infrared plasmonic biosensing with
  graphene,'' Science {\bf 349}, 165 (2015).

\bibitem{Fey63118}
R.~P. Feynman and F.~L. \mbox{Vernon, Jr.}, \newblock ``The theory of a general
  quantum system interacting with a linear dissipative system,'' Ann. Phys.
  {\bf 24}, 118 (1963).

\bibitem{Tan906676}
Y.~Tanimura, \newblock ``Nonperturbative expansion method for a quantum system
  coupled to a harmonic-oscillator bath,'' Phys. Rev. A {\bf 41}, 6676 (1990).

\bibitem{Tan06082001}
Y.~Tanimura, \newblock ``Stochastic Liouville, Langevin, Fokker-Planck, and
  master equation approaches to quantum dissipative systems,'' J. Phys. Soc.
  Jpn. {\bf 75}, 082001 (2006).

\bibitem{Yan04216}
Y.~A. Yan, F.~Yang, Y.~Liu, and J.~S. Shao, \newblock ``Hierarchical approach
  based on stochastic decoupling to dissipative systems,'' Chem. Phys. Lett.
  {\bf 395}, 216 (2004).

\bibitem{Xu05041103}
R.~X. Xu, P.~Cui, X.~Q. Li, Y.~Mo, and Y.~J. Yan, \newblock ``Exact quantum
  master equation via the calculus on path integrals,'' J. Chem. Phys. {\bf
  122}, 041103 (2005).

\bibitem{Red651}
A.~G. Redfield, \newblock ``The theory of relaxation processes,'' Adv. Magn.
  Reson. {\bf 1}, 1 (1965).

\bibitem{Ber71539}
B.~J. Berne, \newblock ``Time-dependent properties of condensed media,'' in
  {\em Physical Chemistry, An Advanced Treatise}, edited by H.~Eyring,
  D.~Henderson, and W.~Jost, volume~8B, pages 539--716, Academic, New York,
  1971.

\bibitem{Haa7398}
F.~Haake, \newblock ``Statistical treatment of open systems by generalized
  master equations,'' in {\em Quantum Statistics in Optics and Solid State
  Physics: Springer Tracts in Modern Physics, Vol.~66}, edited by
  G.~{H\"{o}hler}, pages 98--168, Springer, Berlin, 1973.

\bibitem{Lin76393}
G.~Lindblad, \newblock ``Brownian motion of a quantum harmonic oscillator,''
  Rep. Math. Phys. {\bf 10}, 393 (1976).

\bibitem{Cal83587}
A.~O. Caldeira and A.~J. Leggett, \newblock ``Path integral approach to quantum
  Brownian motion,'' Physica A {\bf 121}, 587 (1983).

\bibitem{Kam92}
N.~G. van Kampen,
\newblock {\em Stochastic Processes in Physics and Chemistry},
\newblock North-Holland, Amsterdam, 1992.

\bibitem{Yan982721}
Y.~J. Yan, \newblock ``Quantum Fokker-Planck theory in a non-Gaussian-Markovian
  medium,'' Phys. Rev. A {\bf 58}, 2721 (1998).

\bibitem{Yan002068}
Y.~J. Yan, F.~Shuang, R.~X. Xu, J.~X. Cheng, X.~Q. Li, C.~Yang, and H.~Y.
  Zhang, \newblock ``Unified approach to the Bloch-Redfield theory and quantum
  Fokker-Planck equations,'' J. Chem. Phys. {\bf 113}, 2068 (2000).

\bibitem{Xu029196}
R.~X. Xu and Y.~J. Yan, \newblock ``Theory of open quantum systems,'' J. Chem.
  Phys. {\bf 116}, 9196 (2002).

\bibitem{Zha987763}
W.~M. Zhang, T.~Meier, V.~Chernyak, and S.~Mukamel, \newblock
  ``Exciton-migration and three-pulse femtosecond optical spectroscopies of
  photosynthetic antenna complexes,'' J. Chem. Phys. {\bf 108}, 7763 (1998).

\bibitem{Jan08101104}
S.~Jang, Y.-C. Cheng, D.~R. Reichman, and J.~D. Eaves, \newblock ``Theory of
  coherent resonance energy transfer,'' J. Chem. Phys. {\bf 129}, 101104
  (2008).

\bibitem{Lai914391}
B.~B. Laird, J.~Budimir, and J.~L. Skinner, \newblock ``Quantum-mechanical
  derivation of the Bloch equations: Beyond the weak-coupling limit,'' J. Chem.
  Phys. {\bf 94}, 4391 (1991).

\bibitem{Jan022705}
S.~Jang, J.~S. Cao, and R.~J. Silbey, \newblock ``Fourth-order quantum master
  equation and its Markovian bath limit,'' J. Chem. Phys. {\bf 116}, 2705
  (2002).

\bibitem{Esp09205303}
M.~Esposito and M.~Galperin, \newblock ``Transport in molecular states
  language: Generalized quantum master equation approach,'' Phys. Rev. B {\bf
  79}, 205303 (2009).

\bibitem{Mav14054112}
M.~G. Mavros and T.~{Van Voorhis}, \newblock ``Resummed memory kernels in
  generalized system-bath master equations,'' J. Chem. Phys. {\bf 141}, 054112
  (2014).

\bibitem{Jin14244111}
J.~S. Jin, J.~Li, Y.~Liu, X.-Q. Li, and Y.~J. Yan, \newblock ``Improved master
  equation approach to quantum transport: From Born to self-consistent Born
  approximation,'' J. Chem. Phys. {\bf 140}, 244111 (2014).

\bibitem{Wei08}
U.~Weiss,
\newblock {\em Quantum Dissipative Systems},
\newblock World Scientific, Singapore, 2008,
\newblock 3rd ed. Series in Modern Condensed Matter Physics, Vol.\ 13.

\bibitem{Kle09}
H.~Kleinert,
\newblock {\em Path Integrals in Quantum Mechanics, Statistics, Polymer
  Physics, and Financial Markets},
\newblock World Scientific, Singapore, 5th edition, 2009.

\bibitem{Yan05187}
Y.~J. Yan and R.~X. Xu, \newblock ``Quantum mechanics of dissipative systems,''
  Annu. Rev. Phys. Chem. {\bf 56}, 187 (2005).

\bibitem{Leg871}
A.~Leggett, S.~Chakravarty, A.~Dorsey, M.~Fisher, A.~Garg, and W.~Zwerger,
  \newblock ``Dynamics of the dissipative two-state system,'' Rev. Mod. Phys.
  {\bf 59}, 1 (1987).

\bibitem{Mei993365}
C.~Meier and D.~J. Tannor, \newblock ``Non-Markovian evolution of the density
  operator in the presence of strong laser fields,'' J. Chem. Phys. {\bf 111},
  3365 (1999).

\bibitem{Tho012991}
M.~Thoss, H.~B. Wang, and W.~H. Miller, \newblock ``Self-consistent hybrid
  approach for complex dystems: Application to the spin-boson model with Debye
  spectral density,'' J. Chem. Phys. {\bf 115}, 2991 (2001).

\bibitem{Kir35300}
J.~G. Kirkwood, \newblock ``Statistical Mechanics of Fluid Mixtures,'' J. Chem.
  Phys. {\bf 3}, 300 (1935).

\bibitem{Zha15024112}
H.~D. Zhang, R.~X. Xu, X.~Zheng, and Y.~J. Yan, \newblock ``Nonperturbative
  spin-boson and spin-spin dynamics and nonlinear Fano interferences: A unified
  dissipaton theory based study,'' J. Chem. Phys. {\bf 142}, 024112 (2015).

\bibitem{Xu151816}
R.~X. Xu, H.~D. Zhang, X.~Zheng, and Y.~J. Yan, \newblock ``Dissipaton equation
  of motion for system-and-bath interference dynamics,'' Sci. China Chem. {\bf
  58}, 1816 (2015),
\newblock Special Issue: Lemin Li Festschrift.

\bibitem{Yan14054105}
Y.~J. Yan, \newblock ``Theory of open quantum systems with bath of electrons
  and phonons and spins: Many-dissipaton density matrixes approach,'' J. Chem.
  Phys. {\bf 140}, 054105 (2014).

\bibitem{Yan16110306}
Y.~J. Yan, J.~S. Jin, R.~X. Xu, and X.~Zheng, \newblock ``Dissipaton equation
  of motion approach to open quantum systems,'' Frontiers Phys. {\bf 11},
  110306 (2016).

\end{thebibliography}

\end{document}